\documentclass{emulateapj}
\usepackage{psfig} 

\newcommand{\sgr}{\mbox{SGR\,J$0418+5729$~}}
\newcommand{\sgrnos}{\mbox{SGR\,J$0418+5729$}}
\newcommand{\degrees}{\ensuremath{^\circ}}

\begin{document}

\title{Discovery of a new Soft Gamma Repeater: \sgr}

\author{A.~J.~van~der~Horst\altaffilmark{1},
V.~Connaughton\altaffilmark{2},
C.~Kouveliotou\altaffilmark{3},
E.~G\"o\u{g}\"u\c{s}\altaffilmark{4},
Y.~Kaneko\altaffilmark{4},
S.~Wachter\altaffilmark{5},
M.~S.~Briggs\altaffilmark{2},
J.~Granot\altaffilmark{6},
E.~Ramirez-Ruiz\altaffilmark{7},
P.~M.~Woods\altaffilmark{8},
R.~L.~Aptekar\altaffilmark{9},
S.~D.~Barthelmy\altaffilmark{10},
J.~R.~Cummings\altaffilmark{10},
M.~H.~Finger\altaffilmark{11},
D.~D.~Frederiks\altaffilmark{9},
N.~Gehrels\altaffilmark{10},
C.~R.~Gelino\altaffilmark{12},
D.~M.~Gelino\altaffilmark{13},
S.~Golenetskii\altaffilmark{9},
K.~Hurley\altaffilmark{14},
H.~A.~Krimm\altaffilmark{10},
E.~P.~Mazets\altaffilmark{9},
J.~E.~McEnery\altaffilmark{10},
C.~A.~Meegan\altaffilmark{11},
P.~P.~Oleynik\altaffilmark{9},
D.~M.~Palmer\altaffilmark{15},
V.~D.~Pal'shin\altaffilmark{9},
A.~Pe'er\altaffilmark{16},
D.~Svinkin\altaffilmark{9},
M.~V.~Ulanov\altaffilmark{9},
M.~van~der~Klis\altaffilmark{17},
A.~von~Kienlin\altaffilmark{18},
A.~L.~Watts\altaffilmark{17}
C.~A.~Wilson-Hodge\altaffilmark{3}
}
\altaffiltext{1}{NASA Postdoctoral Program Fellow, NASA/Marshall Space Flight Center, Huntsville, AL 35805, USA}
\altaffiltext{2}{University of Alabama, Huntsville, CSPAR, Huntsville, AL 35805, USA}
\altaffiltext{3}{Space Science Office, VP62, NASA/Marshall Space Flight Center, Huntsville, AL 35812, USA}
\altaffiltext{4}{Sabanc\i~University, Orhanl\i-Tuzla, \.Istanbul 34956, Turkey}
\altaffiltext{5}{Spitzer Science Center/California Institute of Technology, Pasadena, CA 91125, USA}
\altaffiltext{6}{Centre for Astrophysics Research, University of Hertfordshire, College Lane, Hatfield AL10 9AB, UK}
\altaffiltext{7}{Department of Astronomy and Astrophysics, University of California, Santa Cruz, CA 95064, USA}
\altaffiltext{8}{Dynetics, Inc., 1000 Explorer Boulevard, Huntsville, AL 35806, USA}
\altaffiltext{9}{Ioffe Physical-Technical Institute of the Russian Academy of Sciences, St. Petersburg, 194021, Russian Federation}
\altaffiltext{10}{NASA/Goddard Space Flight Center, Greenbelt, MD 20771, USA}
\altaffiltext{11}{Universities Space Research Associations, NSSTC, Huntsville, AL 35805, USA}
\altaffiltext{12}{Infrared Processing and Analysis Center, MC 100-22, California Institute of Technology, Pasadena, CA 91125, USA}
\altaffiltext{13}{NASA Exoplanet Science Institute, California Institute of Technology, Pasadena, CA 91125, USA}
\altaffiltext{14}{University of California, Space Sciences Laboratory, Berkeley, CA 94720, USA}
\altaffiltext{15}{Los Alamos National Laboratory, P.O. Box 1663, Los Alamos, NM 84545, USA}
\altaffiltext{16}{Space Telescope Science Institute, 3700 San Martin Dr., Baltimore, MD, 21218, USA}
\altaffiltext{17}{Astronomical Institute ÒAnton Pannekoek,Ó University of Amsterdam, PO Box 94249, 1090 GE Amsterdam, The Netherlands}
\altaffiltext{18}{Max-Planck Institut f\"ur extraterrestrische Physik, 85748 Garching, Germany}
\email{Alexander.J.VanDerHorst@nasa.gov}

\begin{abstract}
On 2009 June 5, the Gamma-ray Burst Monitor (GBM) onboard the {\it Fermi} Gamma-ray Space Telescope triggered on two short, and relatively dim bursts with spectral properties similar to Soft Gamma Repeater (SGR) bursts. Independent localizations of the bursts by triangulation with the {\it Konus-RF} and with the {\it Swift} satellite, confirmed their origin from the same, previously unknown, source. The subsequent discovery of X-ray pulsations with the Rossi X-ray Timing Explorer ({\it RXTE}), confirmed the magnetar nature of the new source, \sgrnos. We describe here the {\it Fermi}/GBM observations, the discovery and the localization of this new SGR, and our infrared and {\it Chandra} X-ray observations. We also present a detailed temporal and spectral study of the two GBM bursts. \sgr is the second source discovered in the same region of the sky in the last year, the other one being SGR~J$0501+4516$. Both sources lie in the direction of the galactic anti-center and presumably at the nearby distance of $\sim2$ kpc (assuming they reside in the Perseus arm of our galaxy). The near-threshold GBM detection of bursts from \sgr suggests that there may be more such ``dim'' SGRs throughout our galaxy, possibly exceeding the population of ``bright'' SGRs. Finally, using sample statistics, we conclude that the implications of the new SGR discovery on the number of observable active magnetars in our galaxy at any given time is $\lesssim10$, in agreement with our earlier estimates.
\end{abstract}


\section{Introduction}

In the last decade, observational evidence for neutron stars with extreme surface dipole magnetic fields ($B\sim10^{14}-10^{15}$ G) or ``magnetars'' \citep{duncan92} has steadily grown, with more than 15 magnetar candidates to date. The majority of magnetars are members of two neutron star populations historically known as Soft Gamma Repeaters (SGRs) and Anomalous X-ray Pulsars (AXPs); a couple were previously classified as Isolated Neutron Stars or Compact Central Objects. Although these systems have tangible differences, they are also linked with a multitude of similar properties \citep[for recent reviews, see][]{woods06,mereg08}. Most sources are visible only in X- and low-energy gamma rays; very few have also been detected in the optical and infra-red, while two sources have been observed at radio wavelengths \citep{camilo06,camilo07}. All but two reside in our Milky Way.

The magnetar population has increased very slowly since the discovery in 1979 of sources of repeated soft gamma-ray bursts \citep{mazets79a,mazets79b,golenetskii84,laros86,laros87}, later associated with a new class of astrophysical objects named SGRs \citep{atteia87,laros87,kouv87}; and the confirmation, through bursting episodes, that AXPs were part of the same group in 2002 \citep{gavriil02}. New members are added in the group when (i) they are detected to emit multiple, soft short bursts and (ii) a spin period is found and a spindown rate is measured, which lead to magnetar-like $B$-field estimates. During the 9 years of operation of the Compton Gamma Ray Observatory (CGRO; 1991-2000), we discovered only one new SGR source, SGR~$1627-41$ \citep{kouv98}. In the first $\sim$4 years of operation of the {\it Swift} satellite, no new source was discovered, although several outbursts from known SGRs were recorded \citep{palmer05,israel08,esposito08}. The {\it Fermi} Observatory was successfully launched on 2008, June 11 and the {\it Fermi}/Gamma-ray Burst Monitor (GBM) began normal operations on July 14. During the first 17 months of operation we recorded emission from four SGR sources, of which only one was a known magnetar: SGR~$1806-20$. The other three detections were two brand new sources, SGR~J$0501+4516$, discovered with {\it Swift} and extensively monitored with both {\it Swift} and GBM; \sgrnos, discovered with GBM, {\it Swift} and {\it Konus-RF}; and SGR~J$1550-5418$, a source originally known as AXP~1E$1547.0-5408$ or PSR~J$1550-5418$ \citep{camilo07}.

We present here the discovery of \sgr and its localization by triangulation in \S\ref{sec:obs}, and the precise source localization with the {\it Chandra} X-ray Observatory and our infrared observations in \S\ref{sec:loc}. In section \S\ref{sec:bursts} we describe the properties of the GBM bursts and in \S\ref{sec:sum} we discuss the implications of our discovery.

\section{{\it Fermi}/GBM Observations and Localization by Triangulation}\label{sec:obs}

The {\it Fermi}/GBM has a wide field of view (8~sr) with broad-band energy coverage (8~keV $-$ 40~MeV) and is, therefore, uniquely positioned to detect transient events \citep[for a detailed description, see][]{mee09}. When GBM is triggered, three types of data are accumulated: CTIME Burst, CSPEC Burst, and Time Tagged Event (TTE) data \citep{mee09,kaneko10}. The TTE data consist of time-tagged photon events for an accumulation time of $\sim$330~s, starting $\sim$30 s before the trigger time, with superior temporal resolution of 2~$\mu$s and fine spectral resolution of 128 energy channels.

\begin{figure}
\begin{center}
\begin{minipage}{0.50\columnwidth}
\psfig{width=\columnwidth,figure=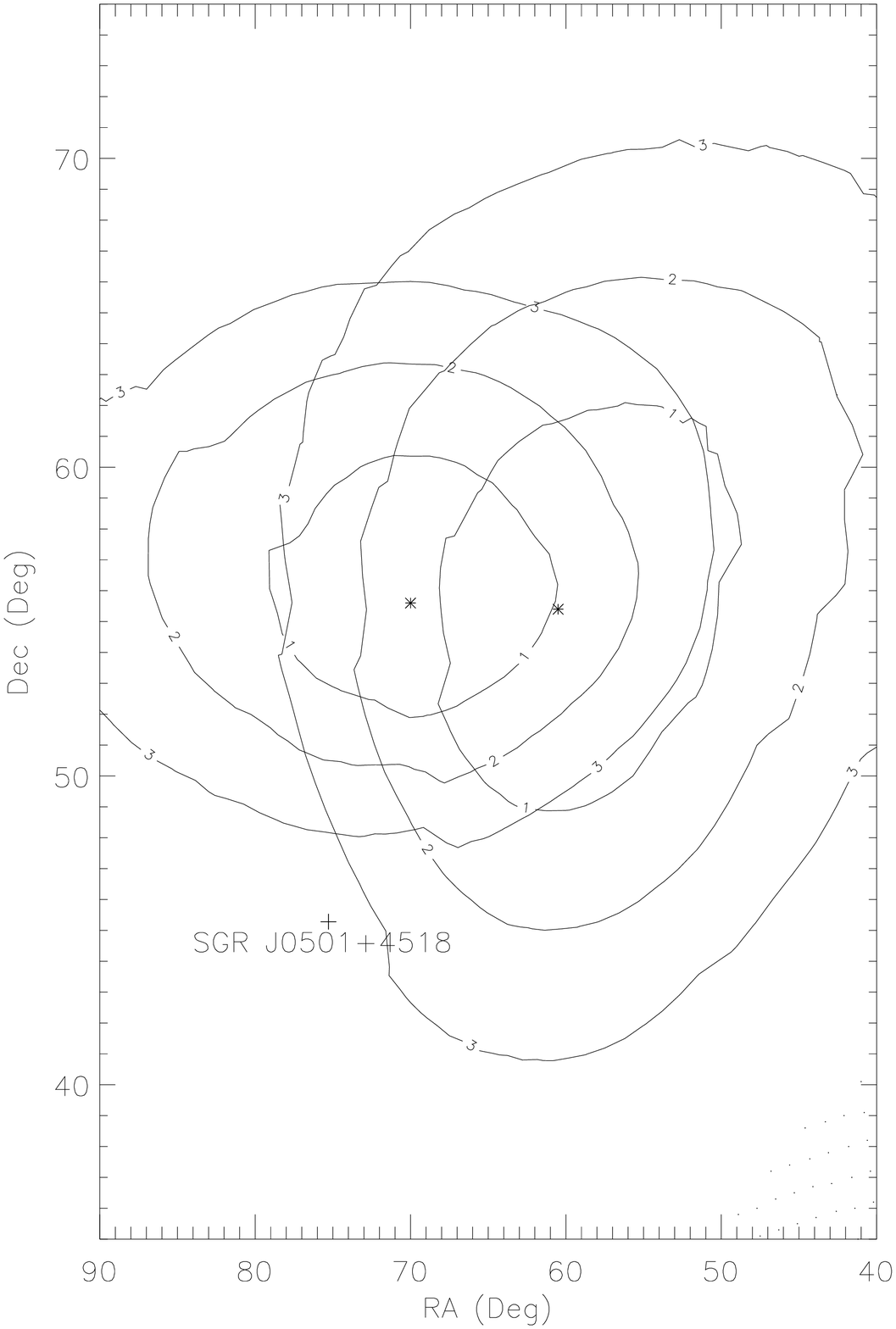}
\end{minipage}
\begin{minipage}{0.46\columnwidth}
\psfig{width=\columnwidth,figure=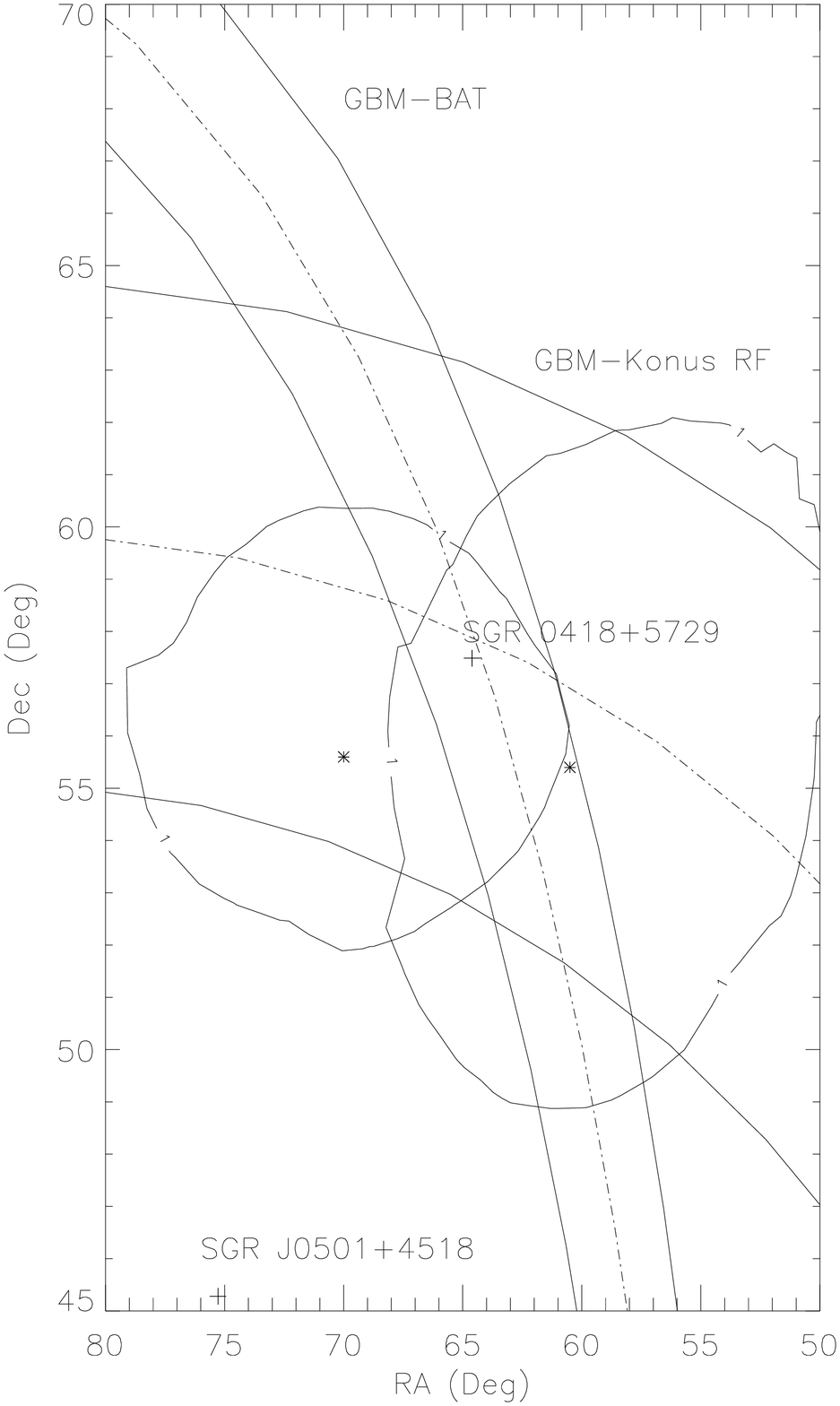}
\end{minipage}
\begin{minipage}{0.75\columnwidth}
\psfig{width=\columnwidth,figure=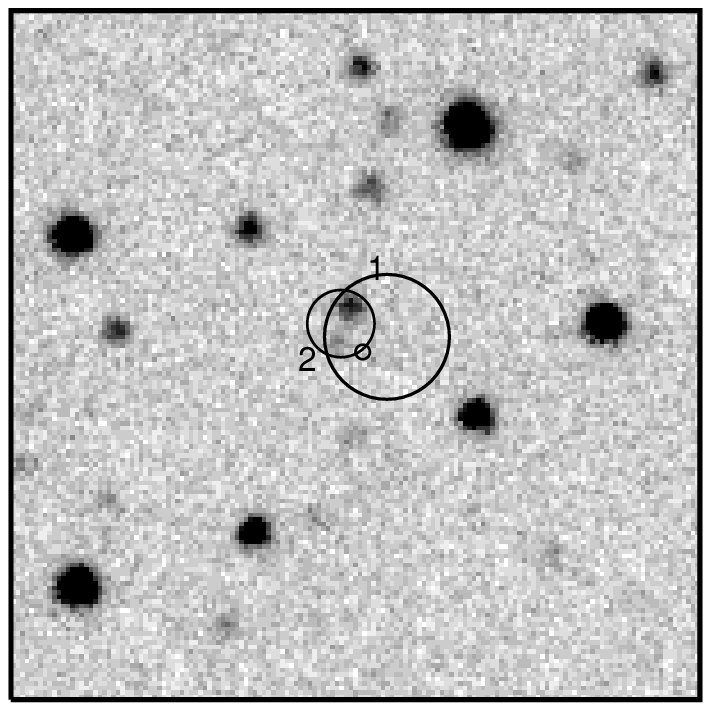}
\end{minipage}
\caption{{\it Top Left:} GBM localizations (asterisks) for the two \sgr triggers on 2009 June 5.  The contours
indicate 1, 2, and 3$\sigma$ statistical uncertainties. The position of the known SGR~J$0501+4516$ is also shown.
It is marginally consistent with the GBM events if one takes into account an additional $2-3\degrees$
systematic component to the localization uncertainties. 
{\it Top Right:} Triangulation annuli (90\% confidence level) for the first \sgr trigger seen by GBM, {\it Konus-RF} and {\it Swift}/BAT on 2009 June 5.  
The asterisks and contours show the $1\sigma$ (statistical uncertainty) localization of both GBM triggers (also shown in top left panel).
A cross marks the  position of the source, found in ground analysis using the {\it Swift}/BAT data (with a $4'$ uncertainty at $90\%$ confidence level).
{\it Bottom:} A $37\arcsec\times37\arcsec$ $K_{\rm s}$ band image of the field of \sgr obtained with Palomar/WIRC. North is up and east to the left. Our {\it Chandra}/HRC error circle with radius $0.35\arcsec$ is shown, as well as the original \citep[$3.6\arcsec$,][]{gogus09b} and refined \citep[$1.9\arcsec$,][]{cummings09} error circles obtained with {\it Swift}/XRT.
}
\label{fig:locs}
\end{center}
\end{figure}

GBM triggered on two SGR-like bursts on 2009 June 5 at 20:40:48.883 UT and 21:01:35.059 UT \citep{avdh09}. 
Their final on-ground calculated locations, $RA, Dec$ (J2000) = 70.0, $+$55.6 (4$^{\rm h}$40$^{\rm m}$, $+55\degrees35\arcmin$) and 
60.5, $+$55.4 (4$^{\rm h}$02$^{\rm m}$, $+55\degrees22\arcmin$), are shown in Figure \ref{fig:locs} (top panels). The positions are consistent at the $1\sigma$ level with a common origin, and inconsistent at the $3\sigma$ confidence level with the known nearby SGR source, SGR~J$0501+4516$, discovered with {\it Swift} in August 2008 (shown in the Figure \ref{fig:locs} top panels). For both triggers, however, there is a systematic component to the localization uncertainty of $2-3\degrees$ so that a reactivation of this known SGR could not initially be excluded, and indeed appeared the most likely origin for these events.

The first GBM burst was seen also by the gamma-ray spectrometer, {\it Konus-RF}, onboard the CORONAS-PHOTON spacecraft and by the {\it Swift}/Burst Alert Telescope (BAT), which was triggered in its partially coded field-of-view.  The second GBM burst was seen weakly in, but did not trigger, the BAT. Triangulation annuli of the GBM-{\it Konus-RF} and the GBM-BAT light curves for the first trigger are shown in Figure \ref{fig:locs} (right upper panel); the GBM localizations are also displayed. Subsequent ground analysis of the BAT data revealed a weak source at $RA, Dec$ (J2000) = 64.606, $+$57.489 (4$^{\rm h}$18$^{\rm m}$25$^{\rm s}$, $+57\degrees29\arcmin16\arcsec$) with an uncertainty of $4\arcmin$, which is consistent with the GBM localizations for both events (right upper panel in Figure \ref{fig:locs}). The annuli clearly exclude the position of SGR~J$0501+4516$; the distance between the two sources is $\sim12\degrees$. Given these results we concluded that GBM detected SGR-like emission from a new source, which we named \sgr \citep{avdh09}. 

\section{Precise Source Location with {\it Swift}, {\it Chandra} and Infrared Observations} \label{sec:loc}

The {\it Swift}/X-Ray Telescope (XRT) observed \sgr starting at 2009 July 8 at 20:52:35 UT (when the source came out of Sun constraints for {\it Swift}) in photon counting mode for a total exposure time of 2.95~ks. A new X-ray source was found at $RA = 04^{\rm h}18^{\rm m}33.70^{\rm s}$, $Dec = +57\degrees 32\arcmin 23.7\arcsec$ (J~2000) with an error circle radius of $3.6\arcsec$. The XRT location is shown in Figure \ref{fig:locs} (bottom panel) and is $\sim3.3\arcmin$ away from the initial {\it Swift}/BAT location. 

To obtain a precise location of the source, we observed the field containing \sgr with the {\it Chandra}/High Resolution Camera (HRC) in imaging mode for 23.8 ks on 2009 July 12. We constructed a binned (by a factor of two) image in the 0.5$-$7 keV band of the entire HRC-I field and searched for point sources at 5$\sigma$ above the background level. We discovered three previously uncatalogued X-ray sources. We searched and detected coherent pulsations at the spin period \citep[9.08~s,][]{gogus09a} of \sgr in the X-ray flux of the brightest X-ray source, which we identified as the X-ray counterpart of \sgr \citep{woods09}. The precise location (absolute positional uncertainty of $0.35\arcsec$ at the 95\% confidence level) of the source is shown in Figure \ref{fig:locs} (bottom panel) and also given in Table \ref{tab:cxo} together with the coordinates of the other two sources.  

The {\it Chandra} position of \sgr was observed with the Wide field Infrared Camera \citep[WIRC,][]{wilson03} on the 5m Palomar Hale telescope on 2009 August 2 \citep{wachter09}. WIRC has a field of view of $8.7\arcmin\times8.7\arcmin$ and a pixel scale of $0.2487\arcsec$/pixel. We obtained 26 $K_{\rm s}$ band images using two 13-position dither scripts (the second script spatially off-set from the first) - each position was a co-added exposure of twelve 5~s images. Atmospheric conditions were very good with seeing $<1\arcsec$ and clear skies during observations.

The individual frames were reduced using a suite of IRAF scripts and FORTRAN programs. These scripts first linearize and dark-subtract the images. A sky frame and flat field image are created from the list of input images, and subtracted from and divided into (respectively) each input image. At this stage, WIRC images still contain a significant bias that is not removed by the flat field. Comparison of 2MASS and WIRC photometric differences across the array shows that this flux bias has a level of $\sim$10\% and the pattern is roughly the same for all filters. Using these 2MASS-WIRC differences for many fields, one can create a flux bias correction image that can be applied to each of the ``reduced'' images. Finally, we astrometrically calibrated the images using 2MASS stars in the field. The images were then mosaicked together and the mosaic was photometrically calibrated using 2MASS stars. Vega magnitudes were computed using the IRAF phot routine with the zero points as found using the 2MASS stars. The final image (see bottom panel of Figure \ref{fig:locs}) has a 5$\sigma$ $K_{\rm s}$ detection limit of 19.6 mag. 

The astrometric solution was derived based on 90 2MASS source matches and carries a formal $1\sigma$ error of $0.1\arcsec$ for the transfer of the 2MASS reference frame to the WIRC image (in addition to the intrinsic $0.1\arcsec$~$1\sigma$ uncertainty of the 2MASS reference system). The two additional X-ray sources (see Table \ref{tab:cxo}) have unambiguous IR counterparts in our WIRC image and hence can be used to tie the X-ray astrometry to that of the IR 2MASS system. We find a small systematic offset between the two reference systems of $0.33\arcsec$ in {\it RA} based on those two sources and no systematic difference in {\it Dec}.  The X-ray positions in Table \ref{tab:cxo} have been corrected for this shift and are thus registered to the 2MASS astrometric reference frame. 

\begin{table}[htbp]
\caption{\small {\it Chandra}/HRC coordinates of \sgr and the two X-ray sources with IR counterparts}
\tabcolsep=2pt
\scriptsize
\begin{center}
\vspace{-2ex}
\begin{tabular}{ccc}
\hline \hline \\[-2ex]
Source & Right Ascension  & Declination  \\
\hline 
 \sgr & 	$04^{\rm h}18^{\rm m}33^{\rm s}.867$ & +57$\arcdeg$32$\arcmin$22.91$\arcsec$ (J2000) \\
\hline									
 CXOU J041819.0+573341 & 	$04^{\rm h}18^{\rm m}19^{\rm s}.097$	 & +57$\arcdeg$33$\arcmin$41.75$\arcsec$ (J2000) \\
 CXOU J041812.5+573154 &  $04^{\rm h}18^{\rm m}12^{\rm s}.578$	 & +57$\arcdeg$31$\arcmin$54.99$\arcsec$ (J2000) \\
 \hline
\end{tabular}
\end{center}
\label{tab:cxo}
\end{table}

Our $K_{\rm s}$ band image overlaid with the X-ray error circles of \citet{gogus09b}, the refined {\it Swift}/XRT position by \citet{cummings09} and our {\it Chandra} position is shown in Figure \ref{fig:locs} (bottom panel). No obvious IR counterpart is detected inside the {\it Chandra}/HRC error circle. Two sources (labelled 1 and 2 in the Figure \ref{fig:locs} bottom panel) with $K_{\rm s}=17.66\pm0.04$ and $K_{\rm s}=18.8\pm0.1$, respectively, are detected within the refined {\it Swift} error circle of \citet{cummings09}. However, our {\it Chandra}/HRC position is sufficiently offset from this {\it Swift}/XRT position to exclude both of these sources. Possibly, a third, very faint source is seen at the southwestern edge of the {\it Chandra}/HRC error circle. Unfortunately, this source is at the detection limit with $K_{\rm s}=21.6\pm1.3$ and cannot be reliably distinguished from a noise spike in the background. Hence our IR observations fail to reveal a convincing counterpart candidate for \sgrnos.

\section{\sgr Burst Analysis} \label{sec:bursts}
 
We have searched the GBM continuous data files for untriggered bursts from \sgr using the algorithm described by \citet{kaneko10} starting two days before the two triggered events and ending six days after. Our search identified only three events from \sgrnos, all detected on 2009 June 5: these include one untriggered burst at 20:35:54.703 UT, and the two triggered events from the source. The untriggered event took place $\sim5$ min before the first GBM trigger and is relatively weak and soft. It was detected only at energies $\lesssim$ 50 keV, and its location is consistent at the $1\sigma$ level with the \sgr location. We also checked the {\it Swift}/BAT data and although the source was in the BAT field-of-view, we did not see a rate increase at the event time.

Additionally, a search through $\gtrsim4000$ IPN events with fluences $\gtrsim7\times10^{-6}$ erg~cm$^{-2}$ and/or peak fluxes $>1$ photon~cm$^{-2}$~s$^{-1}$ in the $25-150$~keV energy range, going back to 1990, does not reveal a significant excess of bursts in the direction of \sgrnos. We conclude that the source did not undergo any episode of intense activity during this time, although we cannot exclude the possibility of isolated, weak events similar to the ones reported in this Letter.

We have performed detailed temporal and spectral analysis on the two triggered events using the TTE data type. The third (untriggered) event was only detected above background as one 256-ms bin in the continuous CTIME data. Since the CTIME data spectral resolution is relatively coarse (only 8 channels), we did not perform a detailed spectral analysis for this event.

The TTE data of the two triggered events were analyzed with the {\it RMFIT (3.2rc2)} spectral analysis software developed for the GBM data analysis\footnotemark{}.\footnotetext{R.S.~Mallozzi, R.D.~Preece, \& M.S.~Briggs, "RMFIT, A Lightcurve and Spectral Analysis Tool," \copyright 2008 Robert~D.~Preece, University of Alabama in Huntsville, 2008} 
We generated Detector Response Matrices using {\it GBMRSP v1.81}. For both events we used in our analysis the three NaI-detectors with the smallest zenith angles (ranging from $26\degrees$ to $44\degrees$) to the source, i.e. NaIs 3, 4 and 5. The $T_{90}$ and $T_{50}$ event durations were estimated in {\it RMFIT} by constructing cumulative fluence plots over the energy range $8-200$ keV for the three detectors combined, and then determining the times during which 90\% and 50\% of the burst counts were accumulated \citep{kouv93}. For the first trigger we find $T_{90} (T_{50})=40\pm7$~ms (10$\pm4$ms), and for the second trigger $T_{90} (T_{50})=80\pm6$~ms (34$\pm4$~ms).

\begin{figure}
\begin{center}
\includegraphics[viewport=0 0 523 720,clip,angle=90,width=\columnwidth]{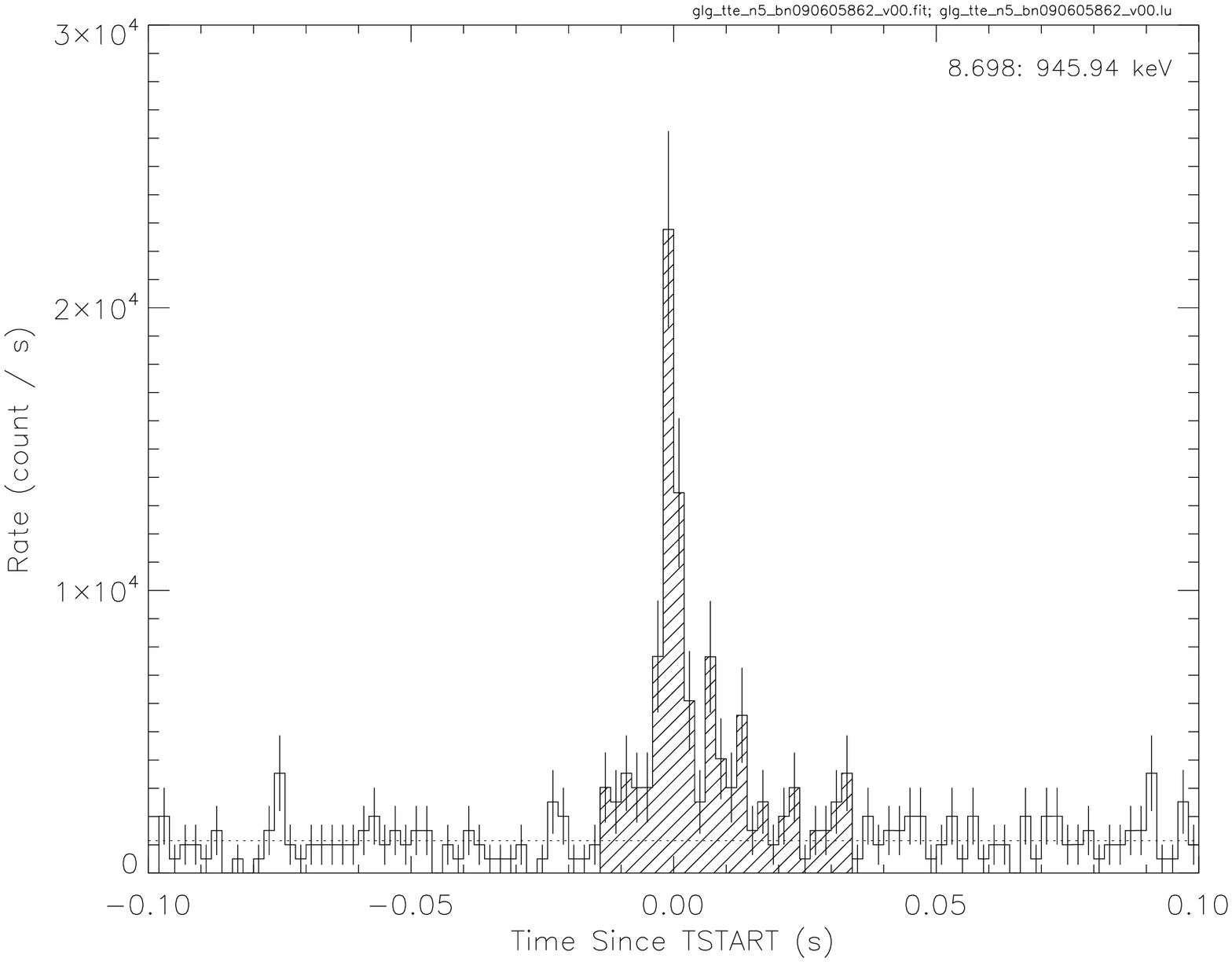}
\includegraphics[viewport=0 0 523 720,clip,angle=90,width=\columnwidth]{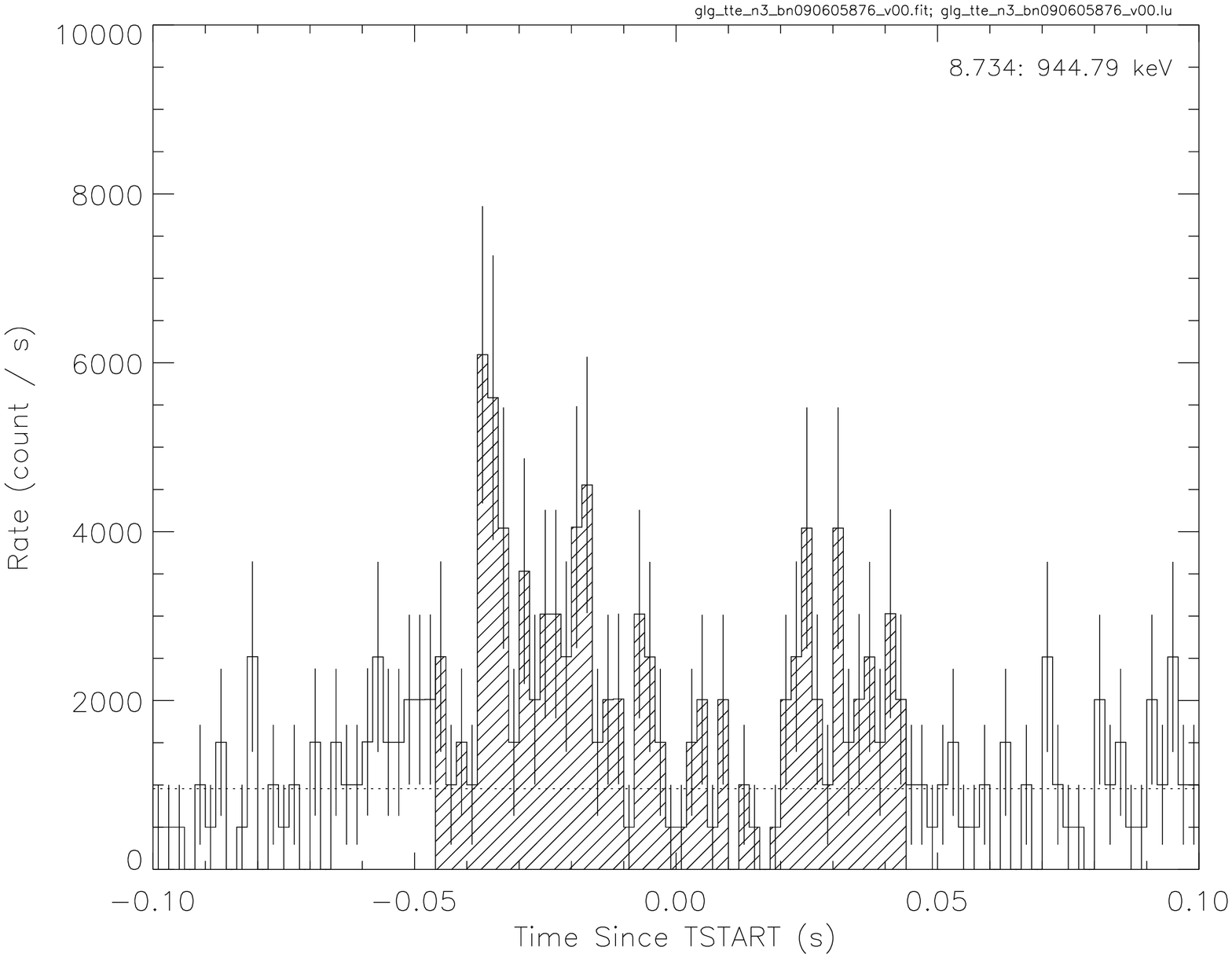}
\caption{TTE light curves of the two \sgr triggered events. 
The hatched areas indicate the time intervals used for spectral analysis.}
\label{fig:lcs}
\end{center}
\end{figure}

\begin{figure}
\begin{center}
\vspace{1ex}
\includegraphics[viewport=0 0 524 760,clip,angle=90,width=\columnwidth]{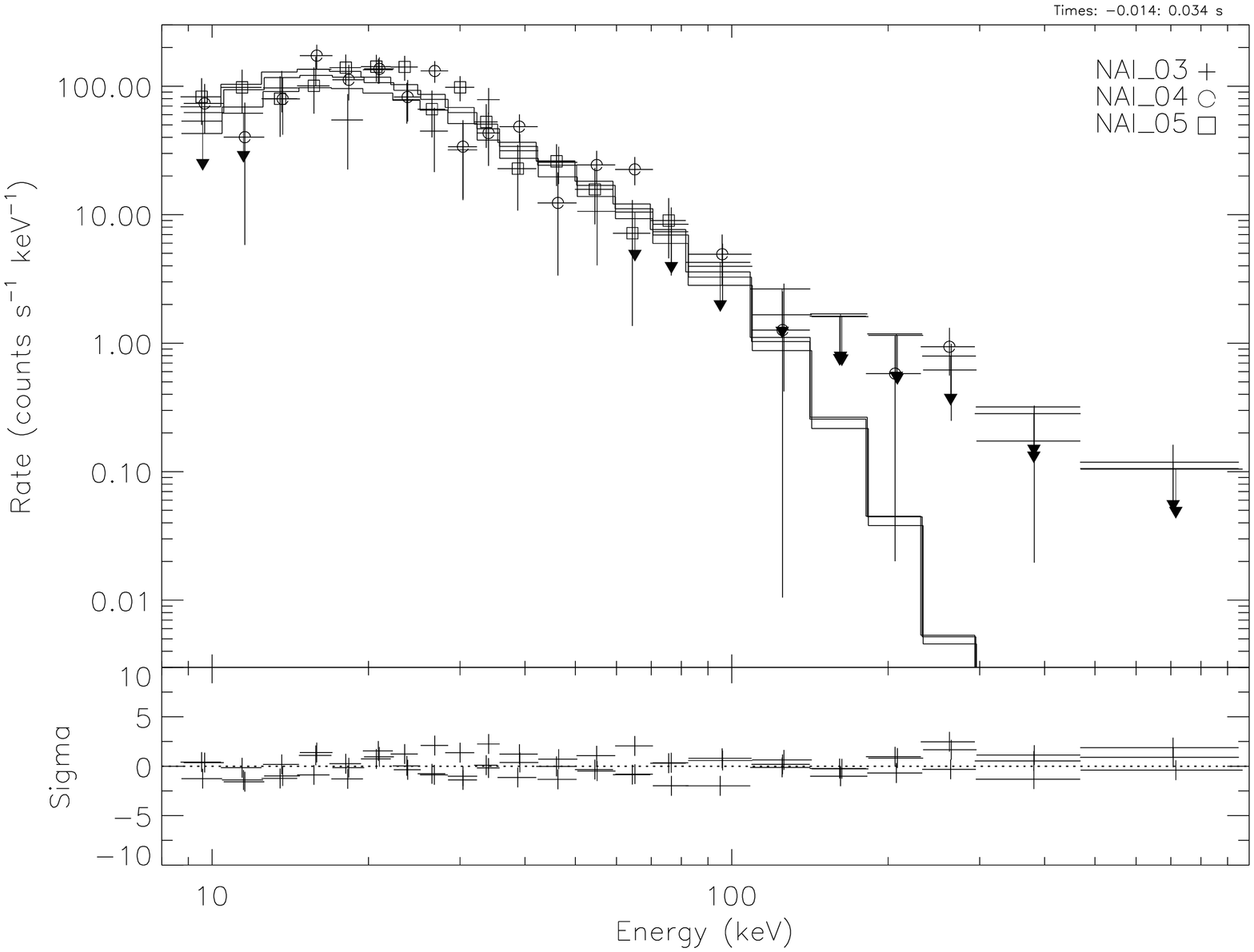}
\includegraphics[viewport=0 0 524 760,clip,angle=90,width=\columnwidth]{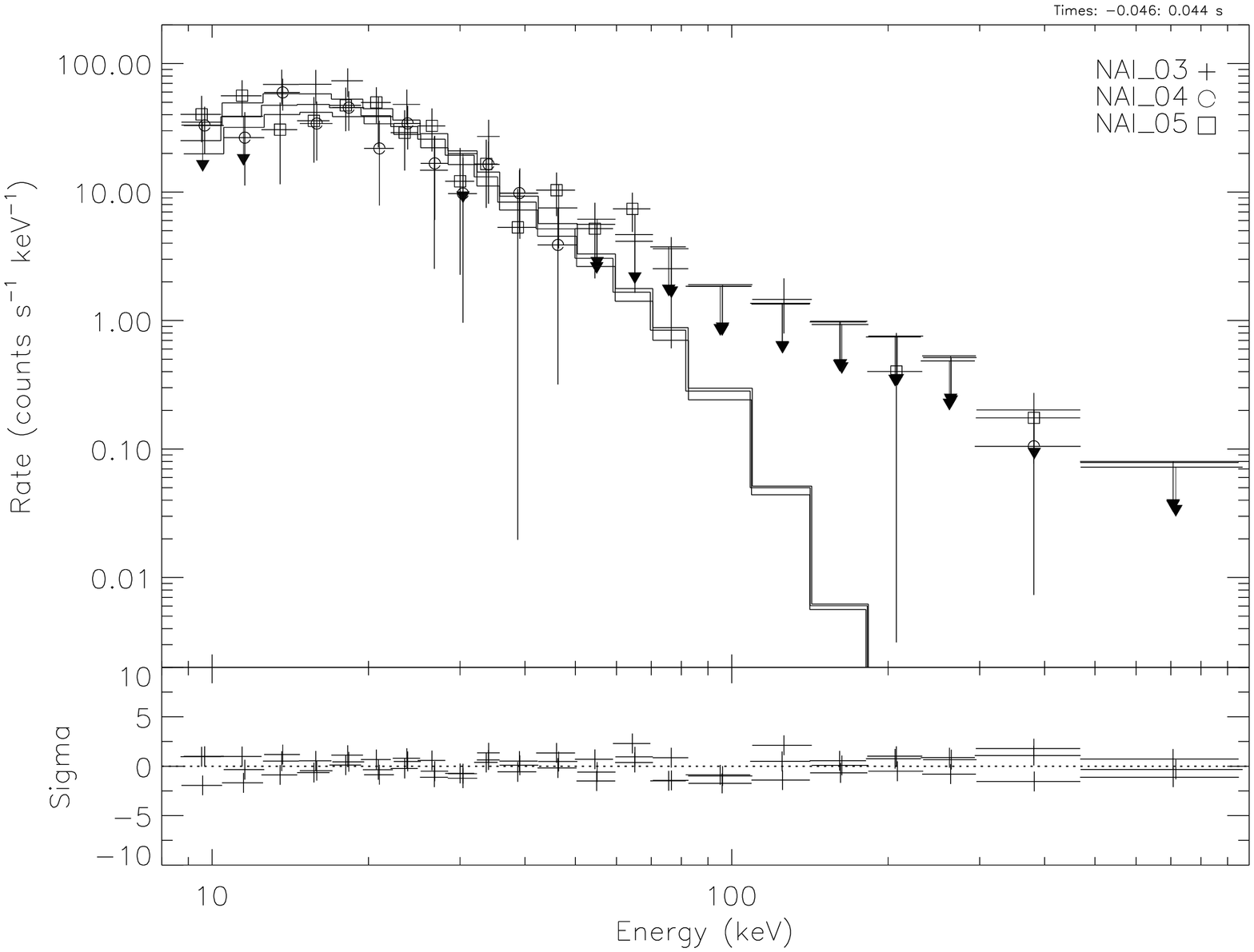}
\caption{Best fit OTTB count spectra of the two \sgr triggered events.}
\label{fig:spectra}
\end{center}
\end{figure}

Figure \ref{fig:lcs} shows the light curve in the detector with the smallest zenith angle for both events, with the intervals used for the spectral analysis indicated. We have fitted the time-integrated spectra with various functions: power law, cut-off power law, black body, and optically-thin thermal bremsstrahlung (OTTB). We find that OTTB provides the best fits in both cases (Table \ref{tab:specresults}) similar to what has been found for other SGR bursts \citep{gogus99,gogus00}. The cut-off power law gives a better statistics value (Table \ref{tab:specresults}), which is not statistically significant given that this model has 1 additional free parameter compared to OTTB. The best-fit count spectra are shown in Figure \ref{fig:spectra}. From the Figure and the fit parameters, it is clear that the spectrum becomes softer from the first to the second burst.

\begin{table*}[htbp]
\caption{\small Spectral analysis results of the \sgr bursts}
\tabcolsep=6pt
\scriptsize \begin{center}
\vspace{-6ex}
\begin{tabular}{ccccccc}
\hline \hline \\[-3ex]
Time & \multicolumn{2}{c}{OTTB} & \multicolumn{3}{c}{Cut-off Power Law} & Energy Flux \\
 & kT & cstat/dof & Index & $E_{\rm peak}$ & cstat/dof & (8$-$200~keV) \\
(UT) & (keV) & & & (keV) & & ($10^{-6}$~erg/cm$^2$/s) \\
\hline
20:40:48.869 $-$ 20:40:48.917 & $33.46\pm 2.23$ & 296.76/361 & $-0.51\pm 0.26$ & $34.72\pm 1.85$ & 294.85/360 & $2.00\pm 0.08$ \\
21:01:35.013 $-$ 20:40:35.103 & $19.71\pm 1.96$ & 335.15/361 & $-0.66\pm 0.52$ & $21.39\pm 2.55$ & 334.85/360 & $0.60\pm 0.04$ \\
\hline
\end{tabular}
\label{tab:specresults}
\end{center}
\end{table*}

To estimate the total energy output of the bursts, we need to know the distance to the source. \sgr is located in the galactic plane and in the galactic anti-center direction. The biggest concentration of stars in that direction is in the Perseus arm of our galaxy, at a distance of $1.95\pm 0.04$~kpc \citep{xu06}. We provide here the \sgr burst energetics, assuming that the source is located in the Perseus arm, which we consider as an upper limit of the energetics. Adopting this distance implies energies (8$-$200~keV) of $\sim 4\times 10^{37}$ and $\sim 2\times 10^{37}$~erg for the two bursts respectively, which is at the lower end of the distribution compared to other SGR bursts \citep{gogus99,gogus00} but at the high end for AXP ones \citep{gavriil04}. We note that an OTTB fit to the CTIME continuous data of the third (untriggered) event, implies an energy (8$-$200~keV) of $\sim 8\times 10^{36}$ erg, albeit with very low statistics.

\section{Discussion}\label{sec:sum}
 
During the first year of operations of the {\it Fermi}/GBM, we have detected bursts from four SGRs, of which two were already known sources and two were newly detected ones. \sgrnos, in particular, was discovered with GBM and located by triangulation with {\it Konus-RF} and {\it Swift}. The source was subsequently confirmed as a magnetar candidate with the Rossi X-ray Timing Explorer ({\it RXTE}) observations \citep{gogus09a}, which revealed a spin period of 9.0783$\pm0.0001$ sec in the persistent X-ray emission of the source. Although the source lies in the direction of the galactic anti-center and presumably at the nearby distance of $\sim2$ kpc (assuming it resides in the Perseus arm of our galaxy), it was not detected at any other wavelength \citep[IR, optical and radio;][]{mignani09,ratti09,lorimer09}. Interestingly, this is the second source discovered in the same region of the sky (including SGR~J$0501+4516$) in one year. The X-ray properties of the source's persistent emission will be described elsewhere (Woods et~al., in preparation).
 
The scarcity of magnetars contributes to the many open questions related to their nature, in particular the true number density and birth rate of these objects. The latest two new SGRs from roughly the same direction, if indeed at the relatively small distance of $\sim$2~kpc, suggest that there are more such ``dim'' SGRs throughout our galaxy, undetectable unless they are relatively close to us. Indeed the trigger detection threshold of the two GBM triggers from \sgr indicates that if their origin was $\sim1.5$ times further away, we would not have detected them. This raises the question: what is the population of such ``dim'' SGRs? A very rough estimate comparing 2 SGRs at $\sim$2~kpc versus 4 at $\sim$10~kpc, and assuming a uniform distribution within the galactic plane, gives $\sim10-15$ times more ``dim'' sources active at a given time. To dominate the magnetar birth rate their active lifetimes should not be larger by more than this factor compared to those of ``bright'' SGRs.

We have estimated the size of the parent population of SGRs using a technique that is commonly employed in the fields of biology and ecology to estimate animal populations \citep{seber82}. This (``Mark and Recapture'') technique is based on capturing and marking a random sample of animals, and then returning them to the population and allowing them to remix.  When a new sample is captured later, the fraction of the recaptured animals that were already marked, and hence are in both samples, can be used to estimate the population size. We applied this technique \citep{seber82} to the SGR observations assuming that the SGR population has (i) fixed membership, and (ii) is homogeneous in its bursting characteristics. There are several caveats associated with these assumptions (e.g., if some SGRs are quiescent for many decades, and then start bursting again, then the sample is biased towards a false ``new'' source), but we are using this estimate as a first order approximation. For the initial sample, we used the number of SGRs (5) found by all instruments in the thirty years prior to GBM (note, that there was not always a complete sky coverage during this period and at times there were several large gaps, when no instrument was available to confirm SGR activity). The GBM observations have resampled the SGR population, finding 4 SGRs, 2 old and 2 new. The unbiased form of the Lincoln-Petersen equation \citep{seber82} estimates that the size of the SGR population is $9~(+17.3/-1.6)$. The interpretation is that GBM is finding 50\% old and 50\% new SGRs, suggesting that the previously known sample is about half of the population observable by GBM.

The discovery of \sgr adds a seventh confirmed member in the SGR subgroup of magnetars in the last 30 years. Together with the known AXPs, the total tally is $\sim15$ sources, a very restricted membership club. Given their small numbers, previous magnetar rate estimates \citep{kouv94,gaensler99,gill07,leahy09} concluded that roughly 10\% of neutron stars become magnetars. Our current rate estimates (based on the currently detectable sources) are consistent with the above, and with our earlier suggestion \citep{kouv94} that our galaxy contains at any given time a few active magnetar sources. However, if a dim, largely undetected as yet magnetar population exists, as the GBM detection of \sgr indicates, it might significantly contribute to and increase the magnetar birth rate.

\acknowledgements{\small This publication is part of the GBM/Magnetar Key Project (NASA grant NNH07ZDA001-GLAST, PI: C. Kouveliotou). {\it Chandra} observations were carried out under Observation ID 10168, part of the proposal ``ToO Observations of SGRs'' (NASA grant GO9-0065Z, PI: C.Kouveliotou). We used data products from the 2 Micron All Sky Survey, which is a joint project of the University of Massachusetts and the Infrared Processing and Analysis Center/California Institute of Technology, funded by the National Aeronautics and Space Administration and the National Science Foundation. We thank T. Jarrett (IPAC/Caltech) for use of his WIRC data reduction software. The {\it Konus-RF} experiment is supported by a Russian Space Agency contract and RFBR grant 09-02-12080-ofi\_m. AJvdH was supported by an appointment to the NASA Postdoctoral Program at the MSFC, administered by Oak Ridge Associated Universities through a contract with NASA. YK and EG acknowledge EU FP6 Transfer of Knowledge Project ``Astrophysics of Neutron Stars'' (MTKD-CT-2006-042722). JG gratefully acknowledges a Royal Society Wolfson Research Merit Award. KH is grateful for support under the Swift Guest Investigator program, NASA grant NNX09AO97G.}

\end{document}